\documentclass[%
 reprint,
 prd,
 amsmath,amssymb,
 aps,
 lengthcheck,
floatfix,
]{revtex4}

\usepackage{graphicx}% Include figure files
\usepackage{dcolumn}% Align table columns on decimal point
\usepackage{bm}% bold math
\usepackage{hyperref}% add hypertext capabilities
\usepackage[utf8]{inputenc}
\usepackage{multirow}
\usepackage{soul}
\usepackage{float}
\usepackage{graphicx}
\usepackage{times}
\usepackage[normalem]{ulem}
\usepackage{color}
\usepackage{cellspace}
\usepackage{lipsum}% http://ctan.org/pkg/lipsum
\usepackage{adjustbox} %rotate table
\usepackage{rotating}
\usepackage{tikz}
\usepackage{changepage}

\newcommand{\be}{\begin{equation}}
\newcommand{\ee}{\end{equation}}
\newcommand{\bea}{\begin{eqnarray}}
\newcommand{\eea}{\end{eqnarray}}

\newcommand{\comment}[1]{}
\renewcommand\sout{\bgroup \color{red} \ULdepth=-.5ex \ULset}
\def\simge{\mathrel{\rlap{\raise 0.511ex
     \hbox{$>$}}{\lower 0.511ex \hbox{$\sim$}}}}
\def\simle{\mathrel{\rlap{\raise 0.511ex
      \hbox{$<$}}{\lower 0.511ex \hbox{$\sim$}}}}

\DeclareUnicodeCharacter{2212}{-}
\usepackage[none]{hyphenat}

\begin{document}

% \setcounter{page}{1}
% \vspace*{0.3 true in}

\title{ 
Learning the relations between neutron star and nuclear matter properties with symbolic regression
}

 \author{\href{https://orcid.org/0000-0003-0103-5590}N. K. Patra$^{1}$\includegraphics[scale=0.06]{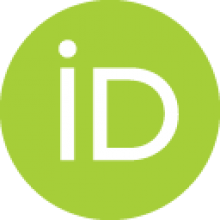}}
 \email{nareshkumarpatra3@gmail.com}
 \author{\href{https://orcid.org/0000-0003-2633-5821}Tuhin Malik$^3$\includegraphics[scale=0.06]{Orcid-ID.png}}
 \email{tm@uc.pt}

 \author{\href{https://orcid.org/0000-0001-9859-1758}Kai Zhou$^{1,2}${\includegraphics[scale=0.06]{Orcid-ID.png}}}\email{zhoukai@cuhk.edu.cn }

 \author{\href{https://orcid.org/0000-0001-6464-8023}Constança Providência$^3$\includegraphics[scale=0.06]{Orcid-ID.png}}
 \email{cp@uc.pt}

\affiliation{$^1$School of Science and Engineering, The Chinese University of Hong Kong (Shenzhen), Shenzhen, Guangdong, 518172, China}
\affiliation{$^2$School of Artificial Intelligence, The Chinese University of Hong Kong (Shenzhen), Shenzhen, Guangdong, 518172, China}
\affiliation{$^3$CFisUC, Department of Physics, University of Coimbra, P-3004 - 516  Coimbra, Portugal}

% \author{Collaborators: N. K. Patra, Tuhin Malik, and Constança Providência}

\date{\today}

\begin{abstract}
The equation of state (EOS) of dense matter in neutron stars (NSs) remains uncertain, particularly at supra-nuclear densities where complex nuclear interactions and the potential presence of exotic matter, like hyperons, come into play. 
The complex relationships existing between nuclear matter and neutron star properties are investigated. The focus is on their nonlinearities and interdependencies. In our analysis, we apply a machine learning algorithm known as symbolic regression, paired with principal component analysis, to datasets generated from Bayesian inference over relativistic mean-field models. A systematic Principal Component Analysis has allowed to break down the percentage contribution of each element or feature in the relationships obtained.
This study examines two main models (datasets): the NL model, which includes nucleonic degrees of freedom; and the NL-hyp model, which includes hyperons in addition to nucleons. Our analysis confirms a robust correlation between the tidal deformability of a 1.4 \(M_\odot\) neutron star and $\beta$-equilibrium pressure at twice the nuclear saturation density. {This correlation remains} once hyperons are included. The contribution of the different nuclear matter properties at saturation to the radius and tidal deformability was calculated. It was shown that the isovector properties have the largest impact, with a contribution of about 90\%. We also studied the relationship between the proton fraction at different densities and various symmetry energy parameters defined at saturation density.  For the hyperon data set, we took into account the effects of the negatively charged hyperon $\Xi$ in order to recover the relationships.  Our study reveals the individual impact of various symmetry energy parameters on proton fractions at different densities. Our research suggests that precise measurements of particular neutron star properties in the future could provide valuable insight into the equation of state, particularly at densities two or three times greater than the saturation density.
\end{abstract}

%\pacs{21.30.Fe, 21.65.Cd, 21.65.Mn, 21.65.Ef}
%\keywords{effective interaction}  

\maketitle
\section{Introduction}
%%%%%% Neutron star . . . . .
Neutron stars (NSs) are mysterious celestial objects in the universe. They act as natural laboratories for studying extremely dense matter \cite{Haensel2007, Glendenning:1997wn, Rezzolla:2018jee}. These stars are the leftover cores of supernova explosions and are incredibly dense — possibly packing more than twice the Sun’s mass into a tiny ball of just 10 km radius \cite{Nicholl:2020mkh, Piekarewicz2017, Ozel:2015fia, Woosley:2016hmi}.
At such high densities, the core of neutron stars may contain exotic particles like hyperons, kaons, pions, or even free quarks, along with normal protons and neutrons. However, figuring out their exact composition is a big challenge for scientists. The only way to know what’s inside is by studying the equation of state (EOS) of dense matter, which links pressure and density \cite{Lattimer:2012nd, Hebeler:2013nza, Lattimer:2021emm, Ferreira:2021pni, Zhang:2018vrx, Malik:2018zcf, OmanaKuttan:2022aml}. The EOS is crucial for understanding supernovae, neutron star mergers, and other cosmic events \cite{LATTIMER2000121, Steiner:2012rk}.
Recent discoveries, such as binary neutron star mergers and neutron star-black hole mergers, have given new clues about these stars \cite{Abbot2017, Abbot2018, Abbot2019, Miller:2019cac, Riley:2019yda, Miller:2021qha, Riley:2021pdl}. Nowadays, astrophysicists are increasingly exploring the relationship between nuclear matter properties, NS composition and NS observables, leveraging advanced computational methods including machine learning to detect subtle patterns in complex data \cite{Ferreira:2019bny,Krastev:2021reh,Krastev:2023fnh}. These techniques are proving invaluable in uncovering hidden connections among various physical parameters, enhancing our understanding of these extraordinary objects.

%%% Past machine learning work . . .
Machine learning (ML) has become a powerful tool in studying NSs, helping physicists uncover their hidden physics. Neural networks, in particular, have been widely used to analyze data from astrophysical observations and simulations \cite{Cuoco:2020ogp, Whittaker:2022pkd, Ferreira:2021pni, Ferreira:2022nwh,Zhou:2023pti}. For example, Bayesian Neural Networks (BNNs) have successfully estimated the internal composition of NSs by studying their radius and tidal deformability \cite{Carvalho:2023ele}. These models have also helped predict the dense matter EOS using gravitational wave signals and other measurements \cite{Thete:2022eif}. ML has also been used to predict nuclear binding energies from atomic mass data, helping determine the EOS of NS crusts \cite{Anil:2020lch}.  Beyond traditional methods, artificial neural networks have even estimated properties of hypernuclei (exotic particles that may exist in NSs) \cite{Vidana:2022prf}. One of ML’s biggest strengths is its ability to find complex patterns in neutron star data, such as linking mass-radius (M-R) measurements to the underlying EOS \cite{Soma:2022qnv, Soma:2022vbb,Li:2025obt,Imam:2024xqg, Patra:2025xtd, Wouters:2025zju, Fujimoto:2021zas}. By mapping observations to EOS predictions, neural networks provide fast and accurate results, often matching or even outperforming traditional physics-based models while also being much faster with low computational costs \cite{Morawski:2020izm}.

%% Symbolic regression 
Another powerful ML technique, the symbolic regression method (SRM), is helping to unveil hidden analytic patterns in NS data. Unlike other methods, SRM finds simple, human-readable equations that explain complex relationships in the data \cite{Bomarito2021, Zhang2023, Keren2023}. This makes it easier for physicists to understand and interpret the results, especially when studying dense matter inside NSs.
For example, SR has been used to: (i) Derive new formulas for nuclear physics problems \cite{Wadekar:2020oov}, (ii) Rediscover known physics laws (like the Gell-Mann–Okubo formula) from data \cite{Zhang:2022uqk}, (iii) Model astrophysical phenomena, such as how galaxies form \cite{CAMELS:2020cof}. This approach bridges the gap between ML and physics, providing clear mathematical insights instead of just numerical predictions \cite{Udrescu:2019mnk}. As research continues, SR could reveal new hidden connections between the compositions and properties of a NS. This paper investigates the potential of ML, particularly SR, to uncover the hidden correlations between various nuclear matter parameters (NMPs) at different densities, proton/electron/hyperon fractions at different densities, and various stellar properties for different NS masses.

%% correlation analysis
Correlation analysis is a fascinating approach to finding relationships between various quantities \cite{Carlson:2022nfb, Tews:2016jhi, Vidana:2009is, Margueron:2017eqc, Essick:2021kjb, Lim:2019som, Richter:2023zec}. We apply SR to a wide dataset obtained from relativistic mean field (RMF) models constrained through Bayesian inference in conjunction with all the recent known constraints on the dense matter EOS. We are also motivated to test these relations with other exotic degrees of freedom, such as hyperon datasets. In more detail, the motivation for this work is twofold:
(i) to uncover the most robust correlations between NMPs, proton fraction, and star properties. This is important because it can help us better understand the physics of NSs and the EOS of dense nuclear matter. 
(ii) Test these correlations with other exotic degrees of freedom, such as hyperons. This is important because hyperons are thought to play a role in the cores of NSs, and understanding their effects can help us learn about the properties of these objects.

%% The paper is organized 
The paper is organized as follows. In Section \ref{formalism}, we provide a concise overview of the formalism used in the present study. The comprehensive analysis of the results and the subsequent discussions are presented in Section \ref{results}. Finally, we provide a brief overview of the results in Section \ref{summary}. 

\section{Methodology} \label{formalism}

In the present  section, we present  a brief review of the framework used to generate the datasets that will be considered in our study, followed by a discussion of the main properties of these datasets.  Next, we introduce the algorithm Python Symbolic Regression (PYSR), which will be used in our analysis, and the  statistical method  Principal Component Analysis (PCA) used to reduce the dimensionality of our datasets.

\subsection{The equation of state}
In the present work, we use the RMF model outlined in Ref. \cite{Malik:2023mnx} to analyze the nuclear matter EOS. This approach employs a mean-field theory strategy that encompasses non-linear meson terms, including both self-interactions and mixed mesonic terms.
The Lagrangian density reads as
\begin{equation}
  \mathcal{L}=   \mathcal{L}_B+ \mathcal{L}_M+ \mathcal{L}_{NL}
\end{equation} 
with
\begin{equation}
\begin{aligned}
\mathcal{L}_{B}=& \bar{\Psi}\Big[\gamma^{\mu}\left(i \partial_{\mu}-g_{\omega B} \omega_{\mu}-%\frac{1}{2}
g_{\varrho B} {\boldsymbol{t}} \cdot \boldsymbol{\varrho}_{\mu}\right) \\
&-\left(m_B-g_{\sigma B} \sigma\right)\Big] \Psi \\
\mathcal{L}_{M}=& \frac{1}{2}\left[\partial_{\mu} \sigma \partial^{\mu} \sigma-m_{\sigma}^{2} \sigma^{2} \right] \\
&-\frac{1}{4} F_{\mu \nu}^{(\omega)} F^{(\omega) \mu \nu} 
+\frac{1}{2}m_{\omega}^{2} \omega_{\mu} \omega^{\mu} \nonumber\\
&-\frac{1}{4} \boldsymbol{F}_{\mu \nu}^{(\varrho)} \cdot \boldsymbol{F}^{(\varrho) \mu \nu} 
+ \frac{1}{2} m_{\varrho}^{2} \boldsymbol{\varrho}_{\mu} \cdot \boldsymbol{\varrho}^{\mu}.\\
    			\mathcal{L}_{NL}=&-\frac{1}{3} b ~m ~g_\sigma^3 \sigma^{3}-\frac{1}{4} c g_\sigma^4 \sigma^{4}+\frac{\xi}{4!} g_{\omega}^4 (\omega_{\mu}\omega^{\mu})^{2} \nonumber\\&+\Lambda_{\omega}g_{\varrho}^{2}\boldsymbol{\varrho}_{\mu} \cdot \boldsymbol{\varrho}^{\mu} g_{\omega}^{2}\omega_{\mu}\omega^{\mu},
\end{aligned}
\label{lagrangian}
\end{equation}
Within this context, the $\Psi$ is a Dirac spinor that describes the baryons with mass $m_B$: the nucleons of model NL and the nucleons plus hyperons of  model NL-hyp. 
The {strength of the} interactions that occur between the baryons and the meson fields $\sigma$, $\omega$, and $\varrho$, respectively, with  masses $m_\sigma$, $m_\omega$, and $m_\varrho$, are denoted by the respective couplings $g_{\sigma B}$, $g_{\omega B}$, and $g_{\varrho B}$.  The values of parameters $b$, $c$, $\xi$, and $\Lambda_\omega$ are established in tandem with the couplings $g_i$ (where $i=\sigma, \omega, \varrho$) by enforcing a prescribed set of constraints. These parameters are crucial for defining the strength of nonlinear terms.

The energy per particle of nuclear matter  at a given density $\rho$ can be divided into two parts:
(i) The energy per particle of symmetric nuclear matter (SNM), which is represented by $\epsilon(\rho, 0)$, and,
(ii) A second term that includes the symmetry energy  $S(\rho)$  and the asymmetry parameter $\delta$,
\bea
\epsilon(\rho, \delta) &\simeq&  \epsilon(\rho,0)+S(\rho)\delta^2, \label{eq:EOS}
\eea
Here, the isospin asymmetry $\delta$ is $({\rho_n -\rho_p})/{\rho}$, measuring the difference between the number of neutrons and protons in nuclear matter.  
The energy per particle $\epsilon(\rho, \delta)$  can be rewritten using different features of bulk nuclear matter of order $n$ defined at  saturation density ($\rho_0$), as shown in Eqs. (\ref{eq:be} \& \ref{eq:sym}). Specifically, (i) for SNM, the energy per nucleon given by $\mathbf{e} = \epsilon(\rho_0, 0) (n = 0)$, the incompressibility coefficient  $K_0 (n = 2)$, the skewness  $Q_0 (n = 3)$, and the kurtosis  $Z_0 (n = 4)$; (ii) for the  symmetry energy the parameters are  the symmetry energy coefficient at saturation density  ${J_{sym,0}}(n = 0)=S(\rho_0)$, together with  its slope  ${L_{sym,0}} (n = 1)$,  curvature $K_{sym,0} (n = 2)$, skewness $Q_{sym,0} (n = 3)$, and kurtosis $Z_{sym,0} (n = 4)$, defined by the following expressions
\bea
\chi_0^{(n)} &=&3^n \rho_0^n \left(\frac{\partial^n\epsilon(\rho,0)}{\partial \rho^n}\right)_{\rho_0}, n=2,3,4; \label{eq:be}
\eea
\bea
\chi_{\rm sym,0}^{(n)} &=&3^n \rho
_0^n \left(\frac{\partial^n S(\rho)}{\partial \rho^n}\right)_{\rho_0}, n=1,2,3,4 \label{eq:sym}.
\eea
{In the following, we will also consider the nuclear matter properties calculated at densities $2,\, 3,\, 4\, \rho_0$. In these cases the index $0$ is dropped and the density at which the derivatives are calculated are identified (i.e. $\chi^{(n)}(\rho)$ and  $\chi_{\rm sym}^{(n)} (\rho)$).  }
The dataset employed in this analysis uses the RMF model and  includes non-linear mesonic interactions \cite{Malik:2023mnx} as outlined above. The model parameters and results are determined using a Bayesian inference method.  The constraints imposed in Bayesian inference were:  a few lower-order nuclear saturation properties \cite{Malik:2020vwo, Chabanat:1997un, Chabanat:1997qh}, a neutron star's maximum mass greater than 2 $M_\odot$ \cite{Miller:2021qha, Riley:2021pdl}, and the low-density pure neutron matter EOS from an  N$^3$LO calculation in chiral effective field theory \cite{Hebeler:2013nza, Lattimer:2021emm}. 

The dataset NL-hyp includes exotic degrees of freedom, the hyperons $\Lambda$ and $\Xi$, within the same model, as discussed in Ref. \cite{Malik:2023mnx}. 
The choice of the meson-hyperon couplings in \cite{Malik:2023mnx} took into account the SU(6) quark model for vector mesons, and the couplings to the $\sigma$ field were determined from the binding energies of known hypernuclei  \cite{Fortin:2017cvt, Fortin:2017dsj, Fortin:2020qin}.

\subsection{Dataset}\label{dataset}
The two datasets, NL and NL-hyp {(which correspond, respectively, to the datasets Set 0 and Set 0\_hyp of ref. \cite{Malik:2023mnx})},  considered in this study, include the following data: both datasets include  (i)  the saturation density ($\rho_0$), together with the pressure of $\beta$-equilibrium matter, and the saturation properties of nuclear matter  binding energy per nucleon ($\mathbf{e}$), incompressibility ($K$), skewness parameter ($Q$), kurtosis parameter ($Z$), symmetry energy coefficient ({$J_{sym}$)}, its slope parameter ({$L_{sym}$)}, curvature parameter ($K_{sym}$), skewness  ($Q_{sym}$), and kurtosis ($Z_{sym}$) at densities $\rho=\rho_0$ (as defined above) and $\rho \in [2,3,4] \rho_0$, where $\rho_0=0.153$ fm$^{-3}$ {is the saturation density}; (ii)  proton (electron) fraction  $X^p\,(X^e)$ at various densities, ($\rho=[2, 3, 4] \rho_0$); 
(iii) pressure at various densities, ($\rho=[2, 3, 4] \rho_0$);
(iv) properties of NSs, such as the tidal deformability ($\Lambda_M$) and the radius ($R_M$) for mass $M \in [1.2, 1.4, 1.6, 1.8, 2.0] M_\odot$;  (v)  the NL-hyp dataset  contains six additional quantities, the hyperon fractions ($X^\Lambda$ and $X^\Xi$) at densities $\rho=[ 2, 3, 4] \rho_0$.

\subsection{Sampling}\label{sampling}
{Symbolic regression is a machine learning technique that derives interpretable mathematical equations directly from data.} 
Python Symbolic Regression (\textsc{PySR}) is a leading algorithm in this domain, capable of extracting human-readable equations that describe the underlying patterns in datasets. While deep learning models may achieve higher predictive accuracy, their black-box nature often obscures the physical relationships between variables, making \textsc{PySR} a valuable alternative for interpretable discoveries.

In our study, we employ \textsc{PySR} (following Refs.~\cite{Varun2011, Ferreira2019}) to model the connection between NMPs and NS properties. Our SR framework is followed as illustrated in Figure-1 of Ref. \cite{Patra:2025xtd}. It processes datasets containing NL (55 features) and NL-hyp (61 features) configurations those are various nuclear matter parameters and  particle fractions at different densities and NS properties at different masses.  The workflow begins by extracting feature vectors ($\mathbf{X[n]}$) and target variables ($\mathbf{Y[n']}$). From these, a random subset of features ($\mathbf{n1}$) and a single target are selected and fed into the \textsc{PySR} algorithm. 
The effective configuration of the PySR algorithm requires careful optimization of its hyperparameters, such as the number of iterations and the set of permitted mathematical operators. We implement a dual-strategy optimization framework combining (i) Bayesian optimization and (ii) systematic grid search. The optimal symbolic expression is selected by applying both techniques concurrently and evaluating their outputs.
The equation with maximum Pearson correlation coefficient and minimum relative error (RE, in \%) between $\mathbf{X}$ and $\mathbf{Y}$ is identified as the best equation. 
This optimization and evaluation cycle is executed iteratively, on the order of $\sim$ 1/2 million times and identify the most robust mathematical expressions linking our features and targets.

\subsection{Principal Component Analysis}\label{PCA}
In multivariate data analysis, Principal Component Analysis (PCA) is an essential statistical method that is used to reduce the dimensionality of the dataset while retaining most of its variance \cite{Wold1987, Aflalo2017, Al-Sayed:2015voa, shlens2014, Liu:2020ely, Acharya:2021lbe, Ali2023}.  The objectives of PCA include extracting essential information, reducing dimensionality, and examining the composition of observations and variables \cite{Abdi2010}. In the present work, we have considered the NMPs and NS properties of the equations listed in Tables \ref{tab1} \&\ref{tab2} as our variables, often referred to as features,  and the corresponding {\bf Y} as our target variables. PCA validates the significance of these selected variables by computing their Principal Components (PCs), which are new variables in alternative dimensions representing linear combinations that  capture shared variation patterns. The eigenvalues associated with the PCs indicate the amount of captured variance, with the first PC's largest eigenvalue representing the largest variance.

The following steps are the methodology of PCA analysis:
\begin{itemize}
    \item Using the provided data, the covariance matrix is computed. The covariance matrix calculates the coefficients of determination for two variables.
    \item The covariance matrix undergoes eigenvalue decomposition in order to yield the eigenvectors and eigenvalues. 
    \item The eigenvectors represent the principal components, and the eigenvalues represent the proportional variance captured by the principal components.
\end{itemize}

The data matrix \(\bf{X}\) representing \(I\) samples with \(J\) variables is generated as a \(I \times J\) matrix in order to compute the covariance matrix. To create a standardized data matrix, \(\bf{\dot{X}}\), the mean for each column is subtracted and then divided by the standard deviation. The covariance matrix elements are determined as 
\bea
{{\bf{C}_{ij}}} &=& \frac{1}{n} \sum^n_{k=1} {\dot{\rm \bf{X}}_{ik}}{\dot{\rm \bf{X}}_{jk}}, 
\eea
where \(n\) represents all samples, and \(i\) and \(j\) indicate variables or features. The order of eigenvalues indicates the significance of matching eigenvectors, and the correlation matrix directs the determination of eigenvalues and eigenvectors. The factor score matrix is represented by 
\(\bf{F} = \bf{\dot{X}}\bf{V}\). Where \(\bf{V = \dot{X}A}\) is referred to as a factor loading matrix, and matrix $\bf{A}$ contains eigenvectors. The projection matrix \(\bf{F}\) is used to represent the projections of observations onto primary components. The collected variance is reflected in the contribution of each PC to the original variables, which is prioritized according to the order of the corresponding eigenvalues.
                                          
\section{Results}\label{results}

\begin{table*}[!ht]
%\begin{sidewaystable*}
\caption{{Best-fit equations with the highest correlation coefficients derived from the NL dataset using symbolic regression for the proton ($X^p$) and electron ($X^e$) fractions at different densities. Three types of equations are presented for each target variable: (i) using symmetry energy at specific densities ($J_{sym}(2\rho_0)$, $J_{sym}(3\rho_0)$, $J_{sym}(4\rho_0)$), (ii) linear combinations of symmetry energy parameters ($J_{sym,0}$, $L_{sym,0}$, $K_{sym,0}$, $Q_{sym,0}$, $Z_{sym,0}$), and (iii) neutron star radii at various masses ($R_{1.2}$, $R_{1.4}$, $R_{1.8}$). Pearson's correlation coefficients for both NL and NL-hyp datasets and mean absolute percentage error (MAPE) values  for NL are also provided. The bottom section includes relations incorporating the hyperon fraction ($X^{\Xi}$) for improved predictions in the NL-hyp dataset at higher densities.}}
\label{tab:equations}
\label{tab1}
%\begin{adjustwidth}{-1.0cm}{}
\setlength{\tabcolsep}{0.3pt} 
\renewcommand{\arraystretch}{0.8} 
\begin{tabular}{llccc}
\hline
\multirow{2}{*}{$\bf{Y}\phantom{{X^p(4\rho_0)}.}$}      & \multirow{2}{*}{$\bf{Equation}$}                                                                                            & \multicolumn{2}{c}{{ Correlation }}  & \multirow{2}{*}{{MAPE} (\%) }\\ \cline{3-4} 
&    & \textbf{NL}    & \textbf{NL-hyp} &  \\ \hline
 &   &              &              &  \\
\multirow{3}{*}{$X^p(2\rho_0)$}   & $0.0034 \times J_{sym}(2\rho_0)- 0.0682$   & 0.99    & 0.99     &  0.72     \\
      & $0.0051\times J_{sym,0}  +0.0011\times L_{sym,0} +0.0001\times K_{sym,0} - 0.112$
   & 0.99         & 0.99    &  6.07    \\
    & $0.0350 \times R_{1.2}-0.3578$  & 0.83         & 0.88 & 5.98           \\
     &      &              &            &  \\
\multirow{3}{*}{${X^p(3\rho_0)}$} & $0.0032 \times J_{sym}(3\rho_0) - 0.0744$       & 0.99         & 0.75  & 0.30        \\
 & $0.0015\times J_{sym,0} + 0.0014\times L_{sym,0} + 0.0001\times K_{sym,0} + 2.0e^{-5}\times Q_{sym,0} -0.0159$ & 0.98         & 0.72     & 4.18       \\
   & $0.0428 \times R_{1.4} -0.4355$     & 0.75         & 0.55    & 8.28        \\
     &   &              &              &  \\
\multirow{4}{*}{${X^p(4\rho_0)}$} & $0.003 \times J_{sym}(4\rho_0) -0.0760$      & 0.99         & 0.04     & 0.47     \\
     &  $0.0048\times J_{sym,0} + 0.0002\times L_{sym,0} +0.0007\times K_{sym,0} + 7e^{-5}\times Q_{sym,0}$  &  &    \\
     & $+ 4.9e^{-6}\times Z_{sym,0} -0.0876$ & 0.96         & -0.01 & 5.41         \\
     & $0.0307 \times R_{1.8} – 0.2596$    & 0.70         & 0.01     &7.95     \\
      &   &              &              &  \\
 \multirow{3}{*}{$X^e(2\rho_0)$}   & $0.0019 \times J_{sym}(2\rho_0)- 0.028$   & 0.99    & 0.99     &  0.69     \\
      & $0.0042\times J_{sym,0}  +0.0097\times L_{sym,0} +0.0012\times K_{sym,0} - 0.1042$
   & 0.97         & 0.92    &  6.61    \\
    & $0.020 \times R_{1.2}-0.1943$  & 0.83         & 0.88 & 4.83           \\
     &      &              &            &  \\
\multirow{3}{*}{${X^e(3\rho_0)}$} & $0.0017 \times J_{sym}(3\rho_0) - 0.0316$       & 0.99         & 0.79  & 0.26        \\
 & $0.0016\times J_{sym,0} + 0.0014\times L_{sym,0} + 0.0002\times K_{sym,0} + 2.1e^{-5}\times Q_{sym,0} -0.0507$ & 0.97         & 0.71     & 5.03       \\
   & $0.023 \times R_{1.4} -0.229$     & 0.76         & 0.38    & 7.16        \\
     &   &              &              &  \\
\multirow{4}{*}{${X^e(4\rho_0)}$} & $0.0016 \times J_{sym}(4\rho_0) -0.0327$      & 0.99         & 0.76     & 0.39     \\
     &  $0.0018\times J_{sym,0} + 0.0012\times L_{sym,0} +0.0003\times K_{sym,0} + 3.9e^{-5}\times Q_{sym,0}$  &  &    \\ & $+ 1.5e^{-6}\times Z_{sym,0} -0.0350$ & 0.95         & 0.71 & 4.40         \\
     & $0.0164 \times R_{1.8} – 0.130$    & 0.70         & 0.17    &7.13     \\    
    &   &              &            &     \\
                                \hline
 &     $\bf{X^p ~at ~higher ~density}$    &              &           &     \\
                                  \hline
     &   &              &              &  \\ 
     
\multirow{4}{*}{${X^p(4\rho_0)}$} & $0.0018 \times J_{sym}(4\rho_0) + 1.0451 \times X^\Xi(4\rho_0) - 0.0760$    & ...         & 0.94    & 2.25    \\
  & $0.0029\times J_{sym,0} + 0.0006\times L_{sym,0} +0.0002\times K_{sym,0} + 1.8e^{-5}\times Q_{sym,0} + 1.7e^{-6}\times Z_{sym,0}$&   &  &    \\
  &$ + 1.579 \times X^\Xi -0.0876$  & ...         & 0.93  &  4.60       \\
  & $0.0164 \times R_{1.8} + 2.525 \times X^\Xi (4\rho_0) –  0.2596$   & ...         & 0.84      &  3.93     \\  \hline
\end{tabular}
%\end{adjustwidth}
%\end{sidewaystable*}
\end{table*}

As discussed earlier, one aim of this research is to investigate a map of multivariate relations, including nonlinearities between various properties of nuclear matter, the pressure of NS matter, the proton and electron fractions at different densities, and the properties of NS at different masses. We are also interested in gaining further knowledge from the symbolic representation of the fractions of strange particles at varying densities, such as hyperons, which could potentially exist in the inner regions of these NSs. For the purpose of this work, we employ the datasets introduced in subsection \ref{dataset} generated through Bayesian inference, see \cite{Malik:2023mnx}. First, we aim to investigate the dataset, which contains only nucleonic degrees of freedom, referred to as the NL dataset. Next, we will apply the learned relations to the dataset with exotic degrees of freedom,  $\Lambda$ and $\Xi$ hyperons, referred to as the NL-hyp dataset, {to examine to what extent the relation remains valid even in the presence of exotic degrees of freedom}.

Our target vectors are selected as proton (electron) fraction at different densities, denoted as \( \mathbf{Y[X^p(\rho)]} \) (\( \mathbf{Y[X^e(\rho)]} \)), where \( \rho\) is [2, 3, 4] \( \rho_0 \). The feature vectors, \( \mathbf{X} \), encompass all remaining quantities in the NL (NL-hyp) datasets. In every iteration, we randomly sample one quantity from \( \mathbf{Y} \) and four quantities from \( \mathbf{X} \), repeating this process for half a  million iterations. The method for choosing the top equations follows the procedure discussed in Sec.\ref{sampling}.

In Table \ref{tab1} we display the best-fit equations with the highest correlation coefficients  obtained through SR from the NL dataset. The focus is on the target variable Y, which represents the proton and electron fractions across different densities. The Pearson's correlation coefficients for both the NL and NL-hyp datasets are also given, as well as the  mean absolute percentage error (MAPE) values for the NL dataset. The MAPE quantifies the average relative deviation between the predicted and actual quantities and is defined as \( \text{MAPE} = \frac{100}{N} \sum_{i=1}^{N} \left| \frac{A_i - P_i}{A_i} \right| \), where \(N\) is the number of data points, \(A_i\) denotes the actual values, and \(P_i\) denotes the corresponding predicted values.

Table \ref{tab1} clearly shows that {for the NL dataset both} the proton fraction  \( X^p\) {and the electron fraction  \( X^e\) are} strongly correlated with the symmetry energy coefficients ($J_{sym}$) at nuclear densities $\rho \in [2,3,4] {\rho_0}$. {The correlation obtained for the electron fraction is a result of the electric charge neutrality condition. The presence of muons does not affect the correlation.}
The linear combinations of iso-vector parameters at saturation density also demonstrate a strong correlation with the proton fractions. The number of iso-vector parameters increases {as the density increases, showing that the role of the higher order iso-vector  NMP on the NS properties becomes stronger as the density increases}. 
The NS radius at different  masses also correlates with the proton fraction ($X^p$) at different  nuclear densities  $\rho \in [2,3,4]{\rho_0}$, although with a weaker correlation. This correlation reflects the existing correlation between the NS radius and the symmetry energy, as seen in Fig. \ref{fig3}, which will be discussed later. The proton fraction at these densities could potentially be determined by simultaneously measuring the radius of NS masses $M\in [1.2,1.4,1.8]\,M_\odot$, {particularly for the low mass stars when the correlation is stronger}.

However, in the NL-hyp dataset, the above correlations involving  the proton fractions at different densities become  weaker or even totally disappear at 4$\rho_0$. This is particularly  true beyond  2 \( \rho_0 \)  because hyperons set in roughly above this density. After the hyperon onset, charged hyperons also enter the charge neutrality condition. Contrary to what is observed with the proton fraction, even at 3 and 4$\rho_0$ the correlations involving the electron fraction are still quite strong. This reflects the fact that the electron fraction has a more systematic behavior at high densities, showing an overall decrease with density, which is not the case with the proton fraction.

To reconstruct the proton fraction correlations at high densities, such as $4\rho_0$ densities, we also allow both $\Lambda$ and $\Xi^-$ fractions, $X^\Lambda,\, X^\Xi$, at $4\rho_0$ as features in X vectors and train the framework on the NL-hyp dataset. After sampling, we found that the proton fraction $X^p$ at $4\rho_0$ correlates with the symmetry energy if the correlation is calculated, including the $\Xi$ fraction. The $\Lambda$ fraction does not couple because $\Lambda$ is charge neutral. The best-obtained relations are written in the last lines of Table \ref{tab1}.

\begin{figure}[!ht]
\centering
\includegraphics[width=0.52\textwidth]{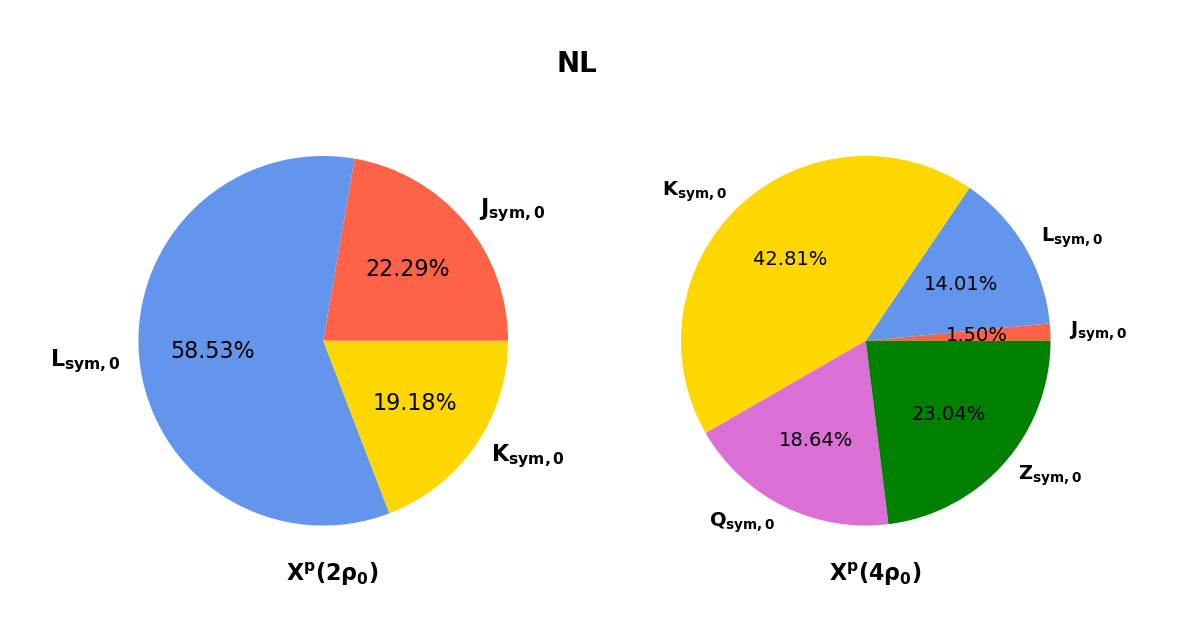}
\includegraphics[width=0.52\textwidth]{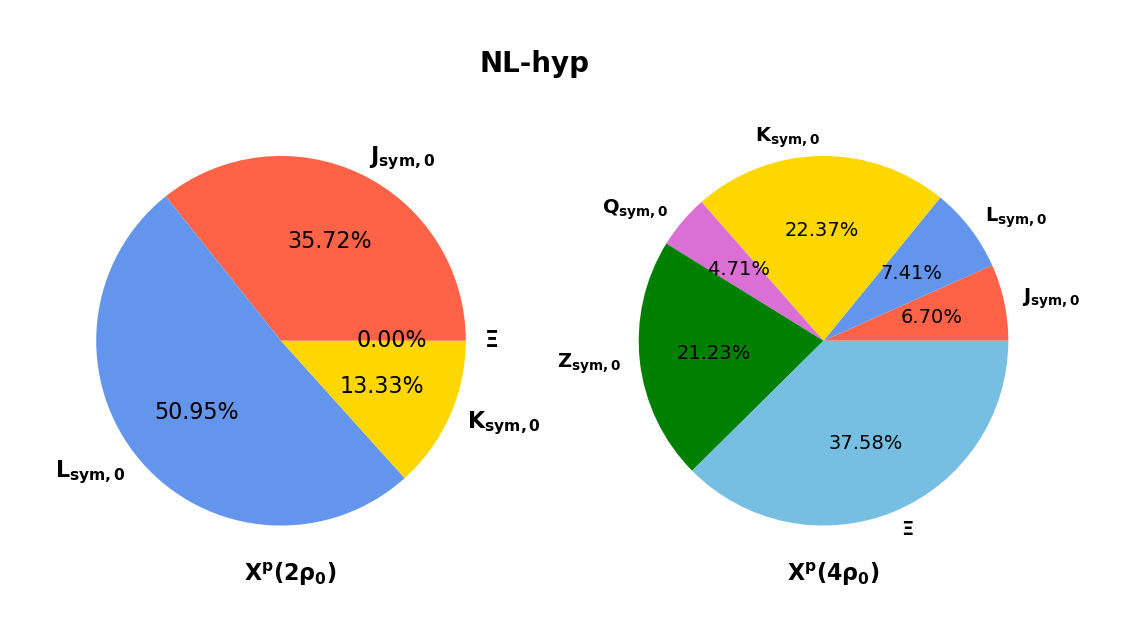}
\caption{The pie chart illustrates the percentage contribution of the saturation properties to the proton fraction ($X^p$) for NL (upper) and NL-hyp (bottom), as indicated in Table \ref{tab1}.}\label{fig1}
\end{figure}

The Fig.~\ref{fig1} displays pie charts that quantify the relative impact of the various nuclear saturation properties on the proton fraction ($X^p$) at two different saturation density levels, specifically at two times ($2\rho_0$) and four times ($4\rho_0$) the normal nuclear saturation density for both the NL dataset (upper panels) and the NL-hyp dataset (lower panels). These contributions are examined using PCA to obtain a more detailed understanding of the equation of state across different densities. We have calculated the contributions of each NMP by considering their impact on the various PCs~\cite{Patra:2023jbz}. We did this by calculating weights from the product of the square of the amplitude for each parameter (features) and the corresponding normalized eigenvalues of PCs. A PC with a normalized eigenvalue of less than 0.1 does not contribute much. The weights were then adjusted so that the total sum of the contributions from all the NMPs equals unity.

For the NL dataset (upper panels), at $2\rho_0$ (upper left), the slope of the symmetry energy ($L_{sym,0}$) shows a predominant impact, accounting for approximately 58.53\% of the contribution, followed by the symmetry energy ($J_{sym,0}$) at 22.29\% and the curvature of the symmetry energy ($K_{sym,0}$) at 19.18\%. At $4\rho_0$ (upper right), $K_{sym,0}$ becomes the dominant contributor at 42.81\%, while the higher-order term $Z_{sym,0}$ contributes 20.04\%, $Q_{sym,0}$ accounts for 18.64\%, and $L_{sym,0}$ decreases to 14.01\%. The contribution from $J_{sym,0}$ becomes minimal at only 1.50\%. These results  reflect the expressions shown in Table \ref{tab1} and are in line with the conclusions of the study~\cite{Alam:2016cli}, where a strong correlation between the radius of low mass stars and a linear combination of the symmetry energy slope and the incompressibility has been found.

For the NL-hyp dataset (lower panels), at $2\rho_0$ (lower left), $L_{sym,0}$ again dominates with a 50.95\% contribution, followed by $J_{sym,0}$ at 35.72\% and $K_{sym,0}$ at 13.33\%. The contributions of the different NMP differ from the those obtained with the NL dataset. This reflects the fact that if hyperons are expected to set in NS and still describe two solar mass stars, NMP properties of nuclear matter have to be adjusted. Notably, the $\Xi^-$ fraction shows no contribution (0.00\%) at this density, as hyperons have not yet appeared.

Conversely, at $4\rho_0$ (lower right), the $\Xi^-$ fraction becomes a major contributor at 37.58\%, reflecting the significant role of hyperons at higher densities. Among the symmetry energy parameters, $K_{sym,0}$ leads with 22.37\%, followed by $Z_{sym,0}$ at 21.23\%, $L_{sym,0}$ at 7.41\%, $J_{sym,0}$ at 6.70\%, and $Q_{sym,0}$ at 4.71\%. This large contribution from the $\Xi^-$ fraction is certainly sensitive to the hyperon couplings.

These charts elucidate how the importance of each saturation property evolves with increasing density. At twice the saturation density, the $L_{sym,0}$ parameter is most influential on the proton fraction in both datasets, indicating its critical role in determining the EOS at lower densities. This indicates that low mass NS may provide valuable information on $L_{sym,0}$. At four times the saturation density, the contributions become more evenly distributed among the higher-order symmetry energy parameters, with $K_{sym,0}$ leading in the NL dataset. However, for the NL-hyp dataset, the appearance of hyperons significantly alters the picture, with the $\Xi^-$ fraction providing the largest contribution. This suggests a shift in the dynamic interplay of nuclear forces at higher densities. Information on the density dependence of the symmetry energy at high densities may be obtained from NS with masses of the order of above 1.4$M_\odot$, although in this case, a possible contribution from hyperons will conceal the behavior of the  symmetry energy.

By leveraging PCA, the complexity inherent in the interactions of nuclear matter at different densities is distilled into a comprehensible visual format, allowing for a clearer interpretation of the nuclear saturation properties' effect on the proton fraction within neutron stars.

Several authors have conducted numerous investigations to limit the range of NMPs based on data about a standard neutron star's radius and tidal deformability. 
The behavior of EOSs near saturation density could have a significant impact on the characteristics of such NSs. However, %the inconsistency in 
the connection between neutron star properties and saturation NMPs is still inconsistent and model dependent \cite{Malik:2018zcf, Alam:2016cli, Carson:2018xri, Forbes:2019xaz, Tsang:2019vxn, Lattimer:2000nx, Guven:2020dok, Reed:2021nqk, Beznogov:2022rri, Tsang:2020lmb}.
In a recent article by {Imam et al.}\cite{Imam:2023ngm}  a direct link between neutron stars' tidal deformability and specific NMPs was obtained through an agnostic approach. By developing a function, they bridge the gap between NS properties and nuclear saturation properties. The research indicates that the tidal deformability of neutron stars with masses between 1.2 and 1.8 $M_\odot$ can be ascertained within a 10\% error using four key nuclear saturation properties. The study by {Patra et al.}\cite{Patra:2023jbz} establishes the connection between neutron star properties and NMPs through a comprehensive multivariate analysis. A PCA is employed as a tool to uncover the connection between multiple NMPs and the tidal deformability as well as the radius of NSs within the mass range of $1.2-1.8M_\odot$.

\begin{table*}[!ht]
\caption{Presented are the optimal equations within the context of minimized features and target variables, where Y exclusively represents the radius ($R_M$) and tidal deformability ($\Lambda_M$) across NS mass 1.2-1.8 $M_\odot$, and the features X consists of NMPs $\in[{\mathbf{e_0}, K_0, Q_0, Z_0, J_{sym,0}, L_{sym,0}, K_{sym,0}, Q_{sym,0}, Z_{sym,0}}]$ at saturation density ($\rho_0$). Pearson's correlation coefficients for the NL and NL-hyp datasets as well as the MAPE values for NL are also provided.}\label{tab2}
\setlength{\tabcolsep}{6.pt}
\renewcommand{\arraystretch}{1.4}
\begin{ruledtabular} 
\begin{tabular}{clccc}
 \multirow{2}{*}{\textbf{Y}} & \multirow{2}{*}{\textbf{Equations}} & \multicolumn{2}{c}{\textbf{Correlation}} & %\multirow{2}{*}
 {\textbf{MAPE }} \\
\cline{3-4}
   &  & \textbf{NL} & \textbf{NL-hyp} &  (in \%)\\
% \cline{4-5}
\hline
{ $R_{1.2}$} & $0.0184\times K_0 + 0.0027\times Q_0 + 0.1460\times L_{sym,0} + 0.0172\times K_{\rm sym,0} + 0.0031\times Q_{\rm sym,0} + 1.309$ & 0.96 & 0.85 & 4.98\\
{ $R_{1.4}$} & $0.0083\times K_0 + 0.0011\times Q_0 + 0.0591\times L_{sym,0} + 0.0080\times K_{\rm sym,0} + 0.0013\times Q_{\rm sym,0} + 7.246$ & 0.97 & 0.90 & 3.69\\
{ $R_{1.6}$}& $0.0083\times K_0 + 0.0009\times Q_0 + 0.0558\times L_{sym,0} + 0.0083\times K_{\rm sym,0} + 0.0012\times Q_{\rm sym,0} + 7.166$ & 0.97 & 0.94 & 4.79\\
{ $R_{1.8}$}& $0.0097\times K_0 + 0.0010\times Q_0 + 0.0665\times L_{sym,0} + 0.0105\times K_{\rm sym,0} + 0.0015\times Q_{\rm sym,0} + 6.148$ & 0.95 & 0.96 & 4.73\\
$\Lambda_{1.2}$ & $2.5840\times K_0 + 0.2141\times Q_0 + 13.659\times L_{sym,0} + 2.1416\times K_{\rm sym,0} + 0.1775\times Q_{\rm sym,0} - 24.89$ & 0.98 & 0.93 & 3.11\\
{ $\Lambda_{1.4}$}& $1.0322\times K_0 + 0.1053\times Q_0 + 5.5840\times L_{sym,0} + 0.9439\times K_{\rm sym,0} + 0.0906\times Q_{\rm sym,0} - 24.30$ & 0.98 & 0.93 & 4.09\\
{ $\Lambda_{1.6}$}& $0.3758\times K_0 + 0.0466\times Q_0 + 2.2189\times L_{sym,0} + 0.3877\times K_{\rm sym,0} + 0.0423\times Q_{\rm sym,0} - 3.611$ & 0.97 & 0.89 & 6.53\\
{ $\Lambda_{1.8}$}& $0.2702\times K_0 + 0.0425\times Q_0 + 1.8834\times L_{sym,0} + 0.3285\times K_{\rm sym,0} + 0.0408\times Q_{\rm sym,0} - 73.98$ & 0.94 & 0.89 & 7.23\\

\end{tabular}
\end{ruledtabular} 
\end{table*}

The compelling findings inspired us to reexamine the relationships between the NS properties such as the dimensionless tidal deformability (\( \Lambda_M \)) and the radius ($R_M$) across various NS masses, and distinct NMPs at saturation density. To achieve this, we utilized SR on the NL dataset, designating \( \mathbf{Y} [\Lambda_M, R_M] \) where $M\in [1.2,1.4,1.6,1.8]M_\odot$ and \( \mathbf{X}[{e_0}, K_0, Q_0, Z_0, J_{sym,0}, L_{sym,0}, K_{sym,0}, Q_{sym,0}, Z_{sym,0}] \) as the target and feature vectors, respectively. For our sampling method, we adhered to the strategy outlined in Sec. \ref{sampling}. However, for each of the 100 iterations, we randomly selected one target from \( \mathbf{Y} \) and four features from \( \mathbf{X} \). It's worth noting that PySR's hyperparameters were optimized to yield the best Pearson's correlation coefficients, as previously undertaken. Table \ref{tab2} details the six most optimal relationships for the radius ($R_M$) and the tidal deformability \( \Lambda_{M} \), where \( M \) ranges between [1.2, 1.4, 1.6, 1.8] \( M_\odot \). Additionally, the table displays the Pearson's correlation coefficients for both the NL and NL-hyp datasets.

Table \ref{tab2}  explains the intricate relationships between NS properties such as radius, tidal deformability, and specific NMPs for neutron stars. Within the confines of minimized features and target variables, the table effectively captures the radius ($R_M$) and tidal deformability (\( \Lambda_M \)) for NS masses ranging from 1.2 to 1.8 \( M_\odot \). It is interesting to note that some NMPs, including \( \mathbf{e_0}, J_{sym,0} \), and \( Z_{sym,0} \), have no effect on both radius and tidal deformability. Overall, all  correlation coefficients decrease in the case of NL-hyp datasets.
The table also illustrates the collaborative role of both iso-scalar and iso-vector components of the EOS in determining the radius and tidal deformability, with the latter  contributing the most. The influence of iso-vector components decreases  with increasing neutron star mass, corroborating findings from prior research by {Patra et al.} \cite{Patra:2023jbz}; {see also \cite{Alam:2016cli, Malik:2018zcf} where individual NMP and linear combinations of two were analyzed. }

\begin{figure*}[!ht]
\centering
\includegraphics[width=\textwidth]{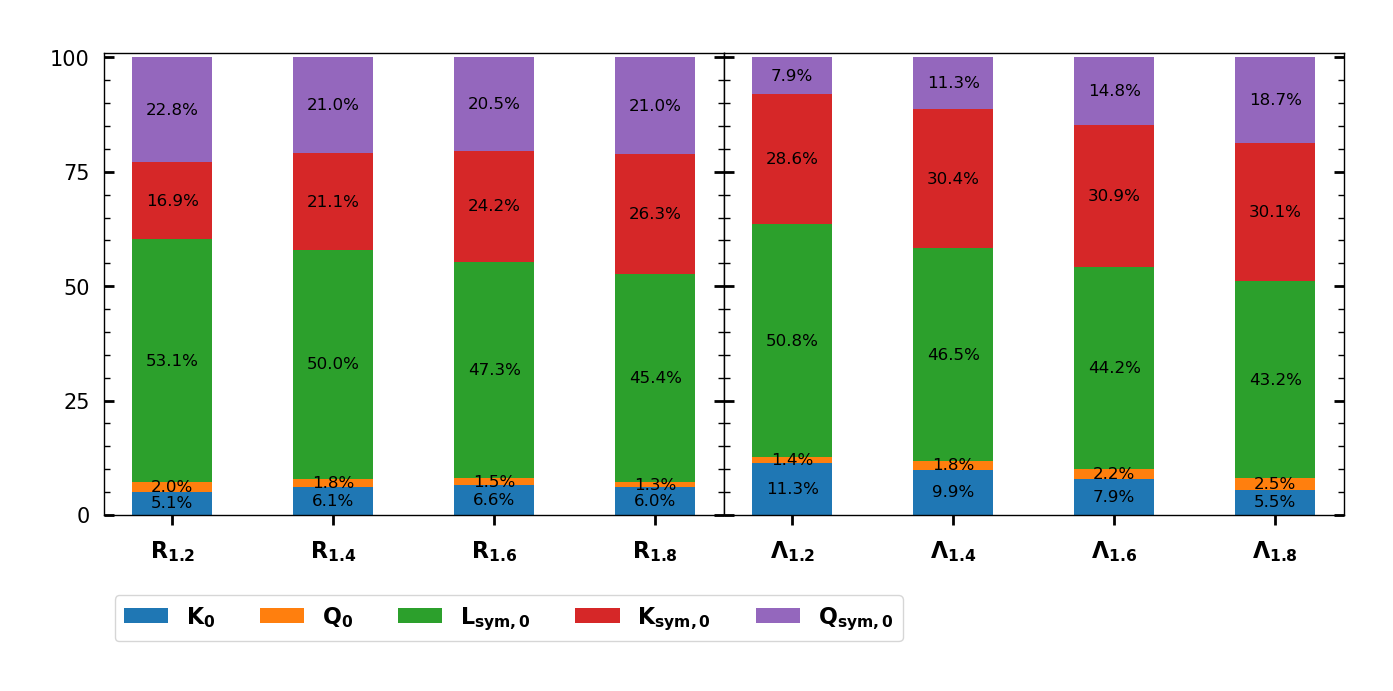}
\caption{The values of the percentage contributions of NMPs to the radius and tidal deformability of NS with masses range $1.2-1.8M_\odot$ (refer to Table \ref{tab2}). %\cp{   Correct $L_0 \to L_{sym,0}$}
}\label{fig2}
\end{figure*}

The Fig. \ref{fig2} illustrates the outcomes of a PCA applied to assess the contributions of various NMPs to the NS properties across different mass ranges, specifically from 1.2 to 1.8 $M_\odot$ (refer to Table \ref{tab2}). Each bar in the graph corresponds to a distinct mass category (1.2, 1.4, 1.6, and 1.8 $M_\odot$) for radius ($R_M$) and tidal deformability ($\Lambda_M$), both significant characteristics of NSs.

The bars are segmented to reflect the percentage contributions from five NMPs at saturation density: the incompressibility ($K_0$), the skewness ($Q_0$), the slope  ($L_{sym,0}$), the curvature  ($K_{sym,0}$), and the skewness  ($Q_{sym,0}$) of the symmetry energy. These parameters are foundational to the equation of state for nuclear matter, which dictates the NS properties.

The slope $L_{sym,0}$ has a dominant influence, above 50\%,  on the radius of low mass  stars with a mass below 1.4$M_\odot$, while $K_{sym,0}$ becomes increasingly significant for higher mass neutron stars and reaches values above 25\% for 1.8$M_\odot$ stars.  $Q_{sym,0}$ and $K_0$  remain approximately constant: they contribute with $\sim 21\%$ and $\sim 6\%$, respectively. The $\Lambda_M$'s sensitivity to $L_{sym,0}$ and $K_{sym,0}$ is substantial across the mass spectrum: the contribution  of $L_{sym,0}$  to  $\Lambda_M$ is slightly smaller than to $R_M$ taking values $\lesssim 50\%$ while the contribution of $K_{sym,0}$ remains approximately constant ($\sim 30\%$). %particularly for 1.8 $M_\odot$, where $K_{sym,0}$ shows a large  contribution.
As the neutron star mass increases from 1.2 to 1.8 $M_\odot$, the relative contribution of $K_0$ decreases to half, while $Q_{\text{sym},0}$ increases to the double of the low mass values, respectively, $\sim 19\%$ and 6\%. 

This visual representation allows us to understand how the significance of each NMP varies with the mass of the NS, providing insights into the complexities of the internal structure of NS and the nuclear forces at play. The PCA consolidates the multidimensional data into PCs, enabling a more manageable interpretation of the EOS parameters' impact on NS properties. {Some important conclusions are:
i) The isovector NMPs primarily determine the NS radius and tidal deformability, with weights of at least 90\%. In contrast, the isoscalar NMP contribution is below 8\% and 13\%, respectively, for the radius and tidal deformability.
ii) The slope $L_{sym,0}$  has the greatest impact of all the main NMPs; however, its contribution is not larger than $\sim$50\%.}

\begin{table}[!ht]
\caption{The optimal equations obtained through feature and target variable minimization. In this context, "Y" specifically denotes the radius and tidal deformability at the canonical NS mass of $1.4M_\odot$, while the features "X" pertain to the $\beta-$equilibrium pressure at densities $\rho \in [2,3,4]\rho_0$. We also provide Pearson's correlation coefficients for the NL and NL-hyp datasets, along with the MAPE values for NL.}\label{tab3}
\setlength{\tabcolsep}{4.pt}
\renewcommand{\arraystretch}{1.4}
\begin{ruledtabular} 
\begin{tabular}{clccc}
 \multirow{2}{*}{\textbf{Y}} & \multirow{2}{*}{\textbf{Equations}} & \multicolumn{2}{c}{\textbf{Correlation}} & \multirow{2}{*}{\textbf{MAPE (in \%)}}\\
\cline{3-4}
   &  & \textbf{NL} & \textbf{NL-hyp} & \\
 %\cline{4-5}
\hline
\multirow{3}{*}{ $R_{1.4}$} & \(0.1118 \times P({2\rho_0}) + 10.18\) & 0.95 & 0.75 & 0.58 \\
& \(0.0304 \times P({3\rho_0}) + 10.31\) & 0.87 & 0.46 & 0.77 \\
& \(0.0118 \times P({4\rho_0}) + 10.54\) & 0.75 & 0.46 & 1.11 \\
\\
\multirow{3}{*}{ $\Lambda_{1.4}$} & \(30.18 \times P({2\rho_0}) - 179.81\) & 0.99 & 0.93 & 1.40 \\
& \(8.01 \times P({3\rho_0}) - 134.33\) & 0.91 & 0.60 & 4.63 \\
& \(3.19 \times P({4\rho_0}) - 65.78\) & 0.77 & 0.45 & 7.40 \\
\end{tabular}
\end{ruledtabular} 
\end{table}

Understanding the behavior of the equations of state in the vicinity of the nuclear saturation density is crucial for characterizing neutron stars. Previous studies have demonstrated a noteworthy correlation between the radius of NSs falling within the mass range of 1-1.4 $M_\odot$ and the pressure of $\beta-$equilibrated matter at densities spanning from 1 to 2 times the nuclear saturation density ($\rho_0$) \cite{Lattimer:2000nx}. This correlation was also confirmed in \cite{Ferreira:2019bgy}, where it was shown that the radius of the NS with larger masses correlates with the pressure at a density above 2$\rho_0$. Similarly, previous studies on tidal deformability have repeatedly demonstrated a significant association between the pressure at twice the nuclear saturation density ($2\rho_0$) and this phenomenon \cite{ Lim:2018bkq, Lim:2019som, Ferreira:2019bgy,Tsang:2019vxn, Tsang:2020lmb,Patra:2022yqc, Imam:2023ngm, Patra:2023jvv}. The persistent trait observed in neutron stars, despite their diverse attributes, is their dependency on the pressure of $\beta-$equilibrated matter at twice the saturation density.

The compelling findings have motivated us to reexamine the correlations between the properties of neutron stars (NS), specifically the dimensionless tidal deformability, (\(\Lambda_M )\) and the radius (\(R_M)\) at the canonical NS mass of 1.4\(M_\odot\), in relation to the pressure at densities within the range of \(\rho \in [2,3,4]\rho_0\). In order to investigate this matter, we have utilized symbolic regression on the NL dataset. The target vectors were designated as \( \mathbf{Y} [\Lambda_{1.4}, R_{1.4}] \), while the feature vectors were denoted as \( \mathbf{X}[P(\rho)] \). The range of \(\rho\) was defined as \([2,3,4]\rho_0\). The sampling procedure employed in our study followed the approach described in Section \ref{sampling}, with the exception, that in each of the 100 iterations, a target was randomly chosen from \( \mathbf{Y} \) and a feature was randomly selected from \( \mathbf{X} \). Similarly to what has been implemented in other studies, the hyper-parameters of PySR were optimized to maximize Pearson's correlation coefficients. 

Table \ref{tab3} displays the six most favorable associations between the $\beta$-equilibrium pressure at 2, 3, and 4$\rho_0$, and  the radius (\(R_{1.4}\)) and tidal deformability (\(\Lambda_{1.4}\)). Furthermore, the table presents Pearson's correlation coefficients for both the NL and NL-hyp datasets. We also found that the radius and the tidal deformability at the canonical NS mass strongly correlate with the pressure at twice the saturation density. The correlation coefficient values decrease as pressure approaches higher densities. This result is consistent with the results obtained in Refs. \cite{Ferreira:2019bgy, Patra:2022yqc}.  {The correlation is much weaker when hyperons are included.}

\begin{figure*}[htp]
\centering
\includegraphics[width=\textwidth]{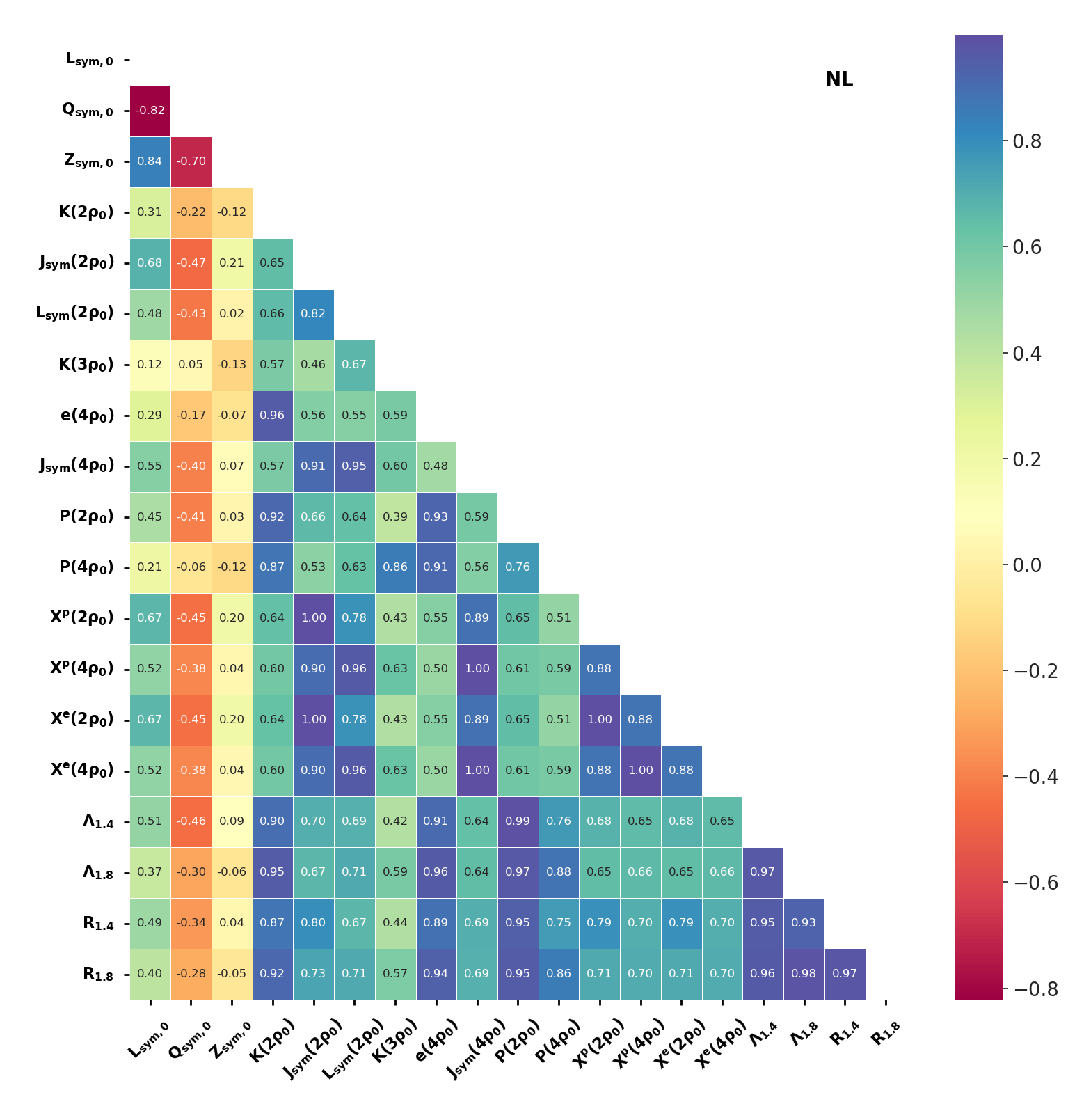}
\caption{The Pearson's correlation coefficients among few selected quantities in NL dataset.}\label{fig3}
\end{figure*} 

\begin{figure*}[htp]
\centering
\includegraphics[width=\textwidth]{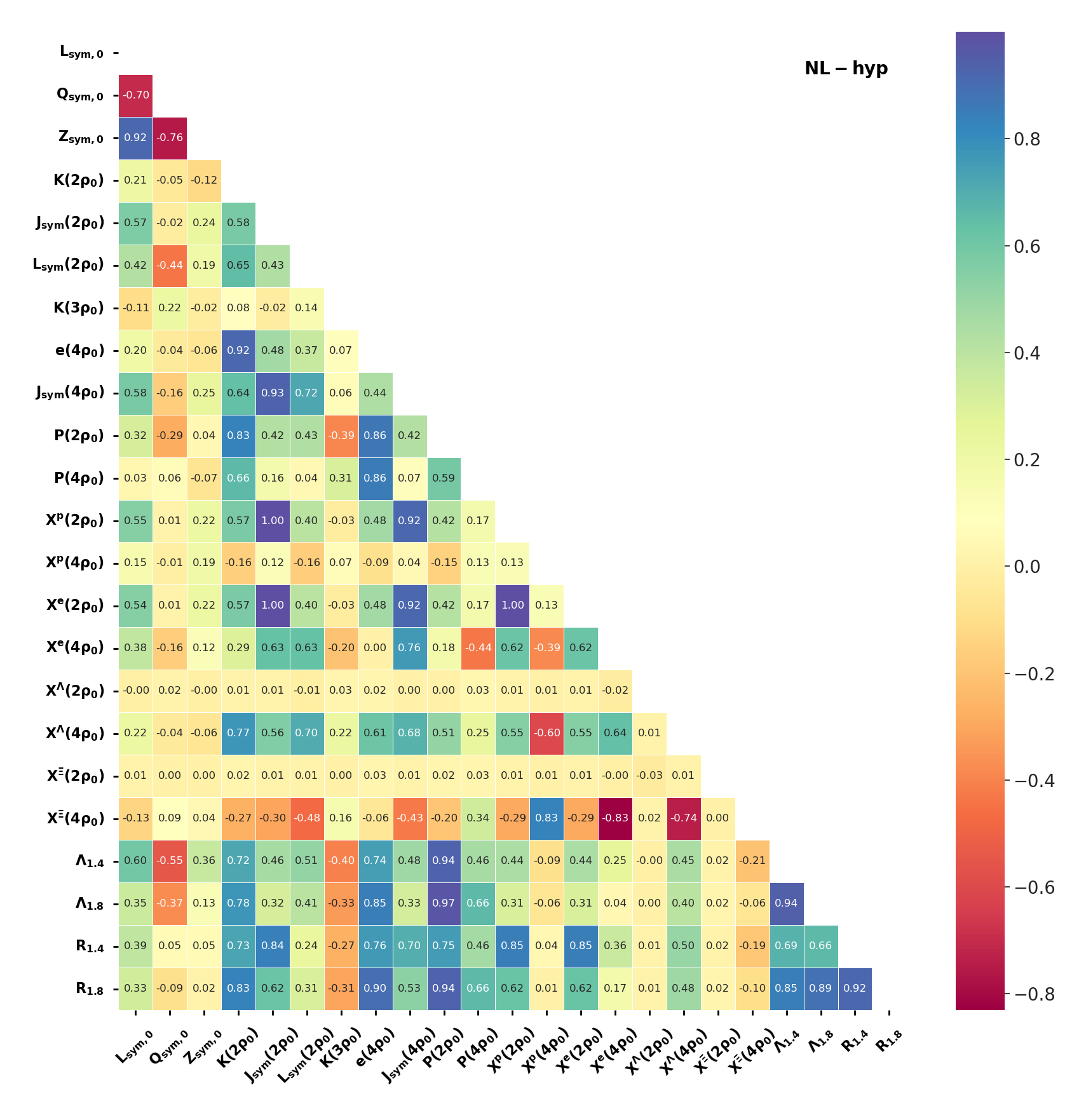}
\caption{Same as Fig.\ref{fig3} but for NL-hyp dataset.}\label{fig4}
\end{figure*}

The Pearson correlation coefficients between several NS properties, NMPs, equation of state ($P(\rho_0)$), and proton (electron) fractions ($X^p(X^e)$) for the NL dataset are displayed in a heatmap in Fig. \ref{fig3}. It is evident that for NSs with masses ranging from 1.4 to 1.8 $M_\odot$, there is a strong relationship with the pressure of $\beta$-equilibrated matter at $2\rho_0$ and isoscalar nuclear saturation properties associated with SNM, in particular, the binding energy per particle $e(4\rho_0)$ and the nuclear incompressibility $K(2\rho_0)$ have
Pearson correlation coefficients of the order of  0.8 or above.  {These correlations reflect the densities found inside NSs with 1.4 and 1.8 $M_\odot$ that lie between 2$\rho_0$ and 4$\rho_0$.}

%There is no significant relationship between the iso-vector nuclear saturation properties associated with the symmetry energy, such as its slope $L_{sym,0}$, $K_{sym,0}$, and $Q_{sym,0}$, and the NS properties such as the radius and tidal deformability. 
The radius and tidal deformability of stars with masses ranging from 1.2 to 1.8 $M_\odot$ are primarily defined by the iso-vector nuclear saturation properties associated with the symmetry energy. However, there is not just one contributing property; see Fig. \ref{fig2}. $L_{sym,0}$ has the largest contribution, but the roles of $K_{sym,0}$ and $Q_{sym,0}$ cannot be neglected. 
However, there is a strong positive correlation, with coefficients  over 0.9, between the proton(electron) fraction ($X^p(X^e)$) at densities $2\rho_0$ and $4\rho_0$ with symmetry energy parameters such as $J_{sym}(2\rho_0)$, $L_{sym}(2\rho_0)$, and  $J_{sym}(4\rho_0)$.

Similarly, in Fig \ref{fig4}, we present the correlation analysis for the  NL-hyp dataset. The results are
strikingly different from the NL dataset. Our findings suggest that due to hyperon fractions, the correlations are reduced in each case, as discussed for NL dataset, considering densities above the hyperon onset. It is found that the correlation of the hyperon fraction at $2\rho_0$ is zero with all quantities because the onset of hyperons occurs above twice the saturation density. The hyperon fractions $X^\Xi(X^\Lambda)$ show somewhat moderate correlations with proton(electron) fractions ($X^p(X^e)$) at $4\rho_0$.

\section{Conclusion}\label{summary}

In this work, we employ symbolic regression, a machine learning technique that derives interpretable mathematical equations directly from data, to investigate the relationships between nuclear matter parameters (NMPs), particle fractions, and neutron star (NS) properties. Our analysis utilizes two datasets generated through Bayesian inference over relativistic mean-field models: the NL dataset containing only nucleonic degrees of freedom and the NL-hyp dataset that additionally includes hyperons ($\Lambda$ and $\Xi^{-}$).

We have confirmed the strong correlation between the radius $R_{1.4}$ and tidal deformability $\Lambda_{1.4}$ with the $\beta$-equilibrium pressure at twice the saturation density, $P(2\rho_0)$, which was previously reported in the literature \cite{Lattimer:2000nx,Lim:2018bkq,Lim:2019som,Ferreira:2019bgy,Tsang:2019vxn}. For the NL dataset, this correlation weakens if the pressure is taken at higher densities. The NL-hyp dataset also shows a strong correlation between $\Lambda_{1.4}$ and $P(2\rho_0)$, while for $R_{1.4}$ this correlation is weaker.

In several studies, the radius and tidal deformability of a  1.4 $M_\odot$ NS  have been used as parameters to limit the range of NMPs. It remains unclear, however, if it is possible to extract NMP  from NS properties due to the complexity of the existing relations.  We have used PySR and a principal component analysis (PCA)  to determine the relation between the radius and tidal deformability of NS masses between 1.2 and 1.8 $M_\odot$ and NMP at saturation. The main conclusions drawn are: i) the iso-vector channel has the largest contribution, at about  90\%. The largest contribution of the  iso-scalar properties $K_0$ and $Q_0$ is of the order of 10\%; ii) the three iso-vector properties that have a larger contribution are $L_{sym,0}$, $K_{sym,0}$ and $Q_{sym,0}$; iii) the contribution of $L_{sym,0}$ is always above 40\%, and even above 50\% for stars with masses below 1.4 $M_\odot$; iv) The contribution of $K_{sym,0}$ to the radius increases with mass, reaching values above 25\% for the largest masses. Meanwhile, its contribution to the tidal deformability remains relatively constant, with values around 30\%. Therefore, we conclude that, although the slope $L_{sym,0}$ contributes significantly to the radii and tidal deformabilities of low-mass stars, its contribution remains at the level of 50\%. The presence of hyperons does not significantly affect the above conclusions.

We also verified that the proton and electron fractions at densities ${\rho \in [2, 3, 4]\rho_0}$ show strong correlations with the symmetry energy $J_{sym}$ evaluated at the corresponding densities. These correlations hold well for the NL dataset but weaken significantly in the NL-hyp dataset at densities beyond $2\rho_0$, when hyperons begin to appear and contribute to charge neutrality. To recover the correlations at higher densities ($4\rho_0$) in the presence of hyperons, we found that including the $\Xi^{-}$ fraction as an additional parameter restores the predictive power of the equations. The $\Lambda$ hyperon fraction does not contribute due to its charge neutrality. 

Through PCA, we quantified the relative contributions of various NMPs to the proton fraction. At $2\rho_0$, the slope of the symmetry energy $L_{sym,0}$ dominates with a contribution above 50\%, while at $4\rho_0$, higher-order parameters such as $K_{sym,0}$, $Q_{sym,0}$, and $Z_{sym,0}$ become increasingly important.

Our results have shown how a symbolic regression approach can help us understand the contributions of different NMPs to NS properties, including radius, tidal deformability and composition, such as particle  fractions. This information is important for understanding the constraints that NSs can impose on the equation of state at supra-saturation densities. We have demonstrated that symbolic regression offers a powerful tool for extracting transparent, physics-informed relationships from complex astrophysical data, bridging the gap between machine learning and traditional nuclear physics approaches.

\section{Acknowledgements} 
NKP would like to acknowledge CFisUC, University of Coimbra, for their hospitality and local support provided during his visit for the purpose of conducting part of this research. KZ acknowledge support by the NSFC grant under No. 92570117, the CUHK-Shenzhen University development fund under grant No. UDF01003041 and UDF03003041, Shenzhen Peacock fund under No. 2023TC0179 and the Ministry of Science and Technology of China under Grant No. 2024YFA1611004. This work was partially supported by national funds from FCT (Fundação para a Ciência e a Tecnologia, I.P, Portugal) under project UID/04564/2025 identified by DOI 10.54499/UIDB/04564/2025. The authors acknowledge the Laboratory for Advanced Computing at the University of Coimbra for providing {HPC} resources that have contributed to the research results reported within this paper, URL: \hyperlink{https://www.uc.pt/lca}{https://www.uc.pt/lca}. 

 %\bibliographystyle{apsrev4-1}
% \bibliographystyle{IEEEtran}
  %\bibliography{ref}

\end{document}